# VR-Stepper: A Do-It-Yourself Game Interface For Locomotion In Virtual Environments

**Denys J.C. Matthies**
University of Munich (LMU)
Geschwister-Scholl-Platz 1
80539 Munich, Germany
matthies@cip.ifi.lmu.de

**Felix M. Manke**
University of Munich (LMU)
Geschwister-Scholl-Platz 1
80539 Munich, Germany
manke@cip.ifi.lmu.de

**Franz Müller**
University of Munich (LMU)
Geschwister-Scholl-Platz 1
80539 Munich, Germany
muellerfr@campus.lmu.de

**Charalampia Makri**
University of Munich (LMU)
Geschwister-Scholl-Platz 1
80539 Munich, Germany
c.makri@campus.lmu.de

**Christoph Anthes**
Leibniz Supercomputing
Center (LRZ)
Boltzmannstr. 1
85748 Garching, Germany
christoph.anthes@lrz.de

**Dieter Kranzlmüller**
University of Munich (LMU)
MNM-Team
Oettingerstr. 67,
80538 Munich, Germany
kranzlmueller@ifi.lmu.de

**ABSTRACT**
Compared to real world tasks, completing tasks in a virtual environment (VE) seldom involves the whole spectrum of skills the human body offers. User input in a VE is commonly accomplished through simple finger gestures, such as walking in a scene by simply pressing a button, even if this kind of interaction is not very suitable. In order to create a more intuitive and natural interaction, diverse projects try to tackle the problem of locomotion in VE's by trying to enable a natural walking movement, which is also supposed to increase the level of immersion. Existing solutions such as treadmills are still expensive and need additional fixation of the body. In this paper, we describe a simple and inexpensive way to build a useful locomotion interface using a conventional sports stepper and an Arduino. This device enables control in a VE by walking-in-place and without the need for any additional fixation gadgets. We conducted a user study with 10 participants to evaluate the impression on the joy and ease of use, immersion and reliability in comparison to other interfaces used for locomotion, such as the Wii Balance Board and a Wand Joystick. We found out that the stepper is experienced slightly better in terms of immersion and joy of use. Furthermore, found that pressing buttons on a Joystick was perceived to be more reliable.

**ACM Classification Keywords**
H.5.2 [Information interfaces and presentation]: User Interfaces—Input devices and strategies, Interaction styles.

**INTRODUCTION**
Technology consistently evolves, thus products as the Oculus Rift[1], which conceptually came from the research field of virtual reality (VR), finally reached a level of being socially acceptable and available to the consumer market. Advertisements for such products and especially computer games show fascinating graphical 3D-environments, which promise the user a totally new experience of VR. In scientific contexts this capturing of senses, diving into an unreal world and feeling a real depth of presence is also called immersion [9]. Nowadays, games and hardware developers often deliver application programming interfaces (APIs) or whole software development kits (SDKs) with their products, so users are enabled to modify and design their own experience. Because these tools became much easier to use, building own 3D virtual worlds doesn't require advanced software engineering skills anymore. Moreover ideas of physical devices can be built more easily. Today, special materials, which used to be available only for research institutes or big industry enterprises, are also accessible for end consumers. Good examples would be Sugru[2], a highly flexible and resistant modeling clay, or low-cost 3D printers[3] for an affordable manufacturing of own product series. Technology has never been that close to the consumer as it is today. Another example for that is the Arduino[4], a board with a microcontroller that allows bringing objects to life. It became very famous because of its simplicity in usage and its large variety of functions. Additional tools such as Fritzing[5] enable an easy learning and understanding of electronics and circuits for not tech-savvy users as well. Beyond this, it is possible to exchange ideas and related technical solutions on the basis of these documentations. The Internet offers a big and growing pool of ideas, hacks and tutorials (e.g. Instructables[6], Hack a Day[7]), which are designed by users for users. Hacking a device and building things on the users own became a great culture, since everything became much easier through the accessible technology[8]. Following these trends, this paper also wants to contribute a do-it-yourself (DIY) solution for a low-cost locomotion interface for virtual environments.

---

[1] Oculus Rift HMD by Oculus VR
*http://www.oculusvr.com* [last access: 03/12/2013]

[2] Sugru by FORMFORMFORM
*http://www.sugru.com* [last access: 03/12/2013]

**RELATED WORK**

**Virtual environments**
One of the main goals of a virtual environment (VE) is to create an as high level of depth of presence as possible, also called immersion [8]. In 1992, Cruz-Neira et al. [3] built a so called CAVE – an audio visual environment, which is basically a small room consisting of display walls, where a virtual 3D scene is displayed. Locomotion in these environments is mostly performed with finger-interaction and tracking of body parts. Physical movement such as natural walking is hardly feasible since space is limited by walls. Also the physical rotation of the user is often problematic since most CAVE-like installations do not have a back screen. Both discoveries have already been stated by Cirio et al. [2]. Therefore, another approach would be to have the display very close to the human eyes, which would mostly be attached to the human head – called head mounted display (HMD). As mentioned, the Oculus Rift became one of the most popular HMD in recent history. So the rotation problem has been overcome, but other problems have appeared, for instance the level of cyber sickness has become an even greater issue, since users are often not able to see a reference point (e.g. their body parts such as their feet) when the display of the HMD surrounds the whole field of view (FOV). Other issues are latency, as stated by LaViola [5]. When being blind for reality, locomotion through physical walking is limited to walking-in-place only. For this kind of setup, gadgets are often installed to hold the user in position, as for the Omni[9]: a locomotion interface that holds the user in ring gadgetry above a slippery bowl to enable free walking. Functions include: walking/running, turning and jumping. One of the main disadvantages is that the user has to learn a special kind of walking. Otherwise, he will be exhausted very quickly. Crouching and strafing are not supported yet.

**Use of Stepper in Research**
Using a stepper as a locomotion interface for VR applications could provide a solution, which was also employed by Wiegand and Brooks [9], who modified a stepper for a military trainings application. Hamano et al. [4] used a stepper for rehabilitation support. Still, the technical solutions demonstrated were expensive and not sufficient. Inspired by works such as the ones from Pausch [7] or Basu et al. [1], we wanted to design a much cheaper and more technically simple solution, which is also usable for any 3D applications.

**VR-STEPPER**
The VR-Stepper is a sports stepper modified with an attached Arduino board to make it usable as an input device for a personal computer. We demonstrate two ways to turn it into an interactive device connected to a computer, with different levels of difficulty for each way.

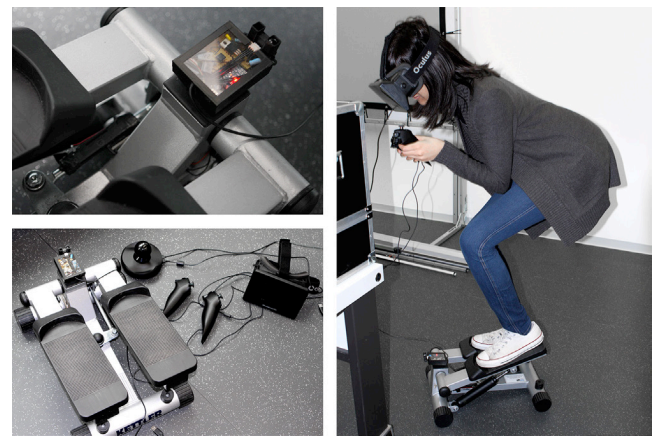

**Figure 1. Crouching User in a VE (Half-Life 2) with the VR-Stepper, Razor Hydra and wearing Oculus Rift.**

---

[3] About.com, List of 3D Printers Under $1000 - Ready To Use, *http://3d.about.com/od/3d-Electronics/tp/List-Of-3d-Printers-Under-1000-Ready-To-Use.htm* [l.a.: 03/12/2013]

[4] Arduino originally by Smart Projects Electronics *http://www.arduino.cc* [last access: 03/12/2013]

[5] Fritzing originally by FH:-P, now: IXDS Berlin *http://www.fritzing.org* [last access: 03/12/2013]

[6] Instructables originally: MIT Media Lab, now: Squid Labs *http://www.instructables.com* [last access: 04/12/2013]

[7] Hack a Day originally by Jason Calacanis *http://www.hackaday.com* [last access: 04/12/2013]

[8] Wettach. R. (2009, November) TED Talk - TEDx Berlin *http://www.tedxberlin.de/tedxberlin-2009-reto-wettach-bodies-secrets* [last access: 03/12/2013]

[9] Omni by Virtuix *http://www.virtuix.com* [last access: 03/12/2013]

**Features of the Stepper**

The VR Stepper enables its user to move in the VE's. A constant motion of the pedals causes a straight *forward* movement, as if the user was really walking. By lowering one pedal until it touches the ground, the user can *turn* around or *strafe* laterally (based on the setup). Additionally, *crouching* on the stepper also results in crouching in a VE.

**Simple DIY Version**

Basic features of the device (walking straight and rotation) can be achieved with the simple DIY version that does not require much knowledge in electronics and no soldering. This minimal version requires the following items:[10]

- Sports Stepper (used: $5-10 new: $39)
- Arduino (Uno: $18 / Nano: $10)
- Potentiometer (slide or turn: $1-2)
- Glue & Tape & Wires (4$)

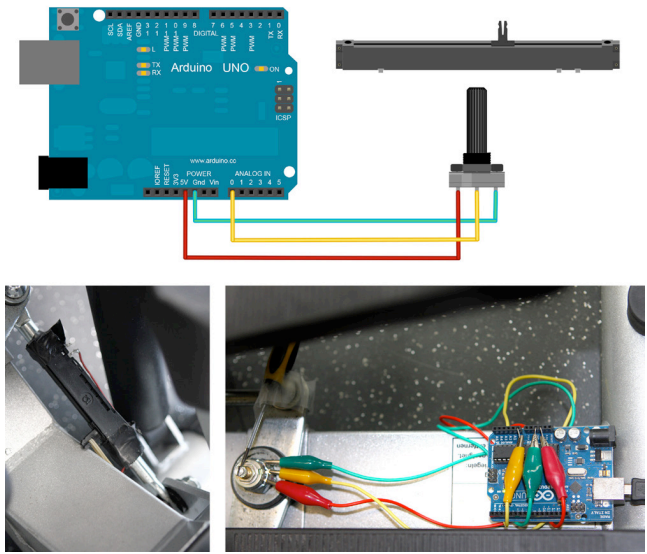

**Figure 2. Finding suitable positions for the potentiometer. A: attached to the rod below the pedal. B: turn potentiometer attached at a pivot to measure the changing angle.**

The main challenge is to attach the potentiometer to a suitable position where the actual motion of the pedals can be measured. Figure 2 demonstrates two suitable positions: (A) a slide potentiometer attached to the rod below the pedal - the knob is bumping against the frame and a rubber band is pulling it back for the countermovement or (B) a turn potentiometer attached at a pivot measures the changing angle - a rod is utilized to fix the knob. Arduino measures the value of the potentiometer and transmits it to the serial port where it is being read out by Processing[11].

---

[10] Average prices checked on ebay.com [15/04/2014].

[11] Processing originally by Ben Fry and Casey Reas (2001) http://processing.org [last access: 11/09/2013]

Lowering the left or right pedal to the ground is shifting or turning the knob of the potentiometer, which should result in a turn or strafe in the VE. Since the value of the potentiometer is being constantly written to the serial port, it can be read out and processed by Processing, where a key-press event can be triggered.

**Advanced Version**

To enable more complex features of the VR-Stepper, namely crouching, a more advanced version has to be built. By building this particular version of the VR-Stepper, one can obtain a device that would be recognized as a keyboard by the computer, making it usable in many programs, such as games where controls are operated by keyboard events with no additional software required. This version needs additional components:

- IR Sensor (Sharp GP2Y0A02YK: $20)
- USB-Keyboard (wired: 5$ wireless: $10)
- 4x Optocoupler ($1)
- 5x 1kOhm Resistor (<$1)
- MOSFET (IRF640: <$2)
- 10uF Capacitor (<$1)
- Universal PCB

*Keyboard Hack*

Unfortunately key events coming from separate software tools are often not suitable for every game, since it could also provide a possibility for cheating. A suitable solution is utilizing the circuit board of a keyboard for transmitting key presses to the computer. These circuits have up to 24 input pins, which create a key press event when connected with each other. Consequently the first thing would be finding out the right keys to utilize (e.g. A, S, D, Directional Arrows, Ctrl key). Because the connections have to be shorten on an electronic way, the pins have to be attached to optocouplers, which have to be connected to the Arduino's output pins. This way the sensor processing must be done at the microcontroller instead of executing a method. Now, the related pins connected to the optocouplers have to be set to HIGH or LOW. Another more expensive but convenient way would be using an *Arduino Leonardo ($30)*, which is able to emulate key press events as well.

*Infrared Sensor*

Additional features as crouching (shown **Figure 1**) would be beneficial to the user as it enlarges the degrees of freedom. Crouching on the stepper is realized with an infrared sensor (*Sharp GP2Y0A02YK*), which has a range of 0-150 cm. This kind of sensor is very simple to interface on the Arduino since it has a behavior similar to a potentiometer: when the user is going down to a crouch position the human body will bend over the IR sensor, thus changing the sensor value to a lower one. Therefore using a threshold is once again a sufficient solution.

*Power Supply*
The keyboard circuit usually gets its power from a USB cable (or a battery) as also the Arduino needs a power supply. It occurs that the power delivered by only one USB port is not sufficient for both devices. Thus having both devices at one USB port a power transistor to amplify the current and a capacitor to smooth the voltage fluctuations are required.

**USER STUDY**
To get some first impressions on the performance of the prototype we conducted a user study. Furthermore we wanted to find out what kind of impact this low-cost locomotion interface has, when applied to a VE.

**Hypothesis**
The first hypothesis (H1) is: moving in a 3D scene by performing leg movements on a Stepper increases immersion. The second hypothesis (H2) would be: letting the user involving their whole body - especially the leg movement - leads to a greater joy compared to common devices such as a Wand Joystick. However, due to the faster and more precise actions possible with a Wand Joystick and having tactile feedback when pushing the button, the Wand Joystick might be rated higher in terms of perceived reliability (H3).

**Pretests**
At first, we decided to use the Stepper as a locomotion interface for playing Half-Life 2[12] with the Oculus Rift and the Razer Hydra[13] (**Figure 1**). However, after two participants, we found out that the users were not able to keep their balance - they fell off the stepper - as soon as the 3D scene was displayed. Therefore the decision was made to use a CAVE-like design, where the users would still have reference points, which should circumvent the problem.

**Study Setup**
The user study was conducted in a stereoscopic 5-display-wall CAVE-like installation (**Figure 3**). To gain knowledge on the performance of the Stepper, we accomplished a within subject study with the 3 following conditions: VR-Stepper, Wii Balance Board[14] and a Wand Joystick[15] in two

---

[12] Half-Life 2 by Valve
*http://orange.half-life2.com* [last access: 05/12/2013]

[13] Razer Hydra Gaming Controller by Razor
*http://www.razerzone.com/gaming-controllers/razer-hydra/* [last access: 05/12/2013]

[14] Wii Balance Board by Nintendo
*http://www.nintendo.com/consumer/downloads/wiiBalanceBoard.pdf* [last access: 06/12/2013]

[15] Flystick 2 by A.R.T.
*http://www.ar-tracking.com/products/interaction-devices/flystick2/* [last access: 06/12/2013]

self-build 3D scenes. The first users' task was it to run as fast as possible through a racing track, which had several hinders, the users had to dodge. The second scene was a waste-land scenario, where the user was enabled for a free walking without any task. Every subject had to go through both scenes by using each interface in a random order. After completing all scenes with every interface a questionnaire had to be filled out, which was asking the user to rate the following on a 5-point Likert scale: (1) Ease of Use (2) Joy of Use (3) Feeling of Immersion (4) Impression on Reliability.

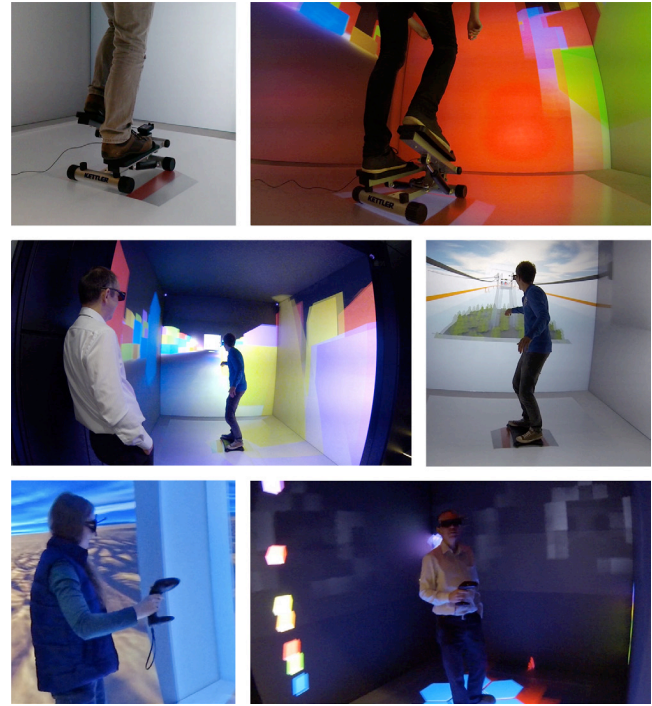

**Figure 3. Experimental setup with the three devices: Stepper, Wii Balance Board & Wand Joystick**

**Participants**
We evaluated the system with one group of 10 participants (8 male, 2 female), with an age between 15 and 55.

**Results**

*Ease of Use*
A one-way ANOVA on the device factor showed a significant difference in terms of ease of use ($F_{2,18} = 5.96$ ; $p=.01$). A Tukey HSD Test suggests that the joystick ($M=3.4$) is significantly easier to use than the VR-Stepper ($M=2.2$ ; $p < .01$). No other differences yield.

*Joy of Use*
A one-way ANOVA showed no significant difference ($p >.05$) between all devices. However, the VR-Stepper ($M=4$) was deemed more joyful than the joystick ($M=3.3$) and the Wii Balance Board ($M=3.5$).

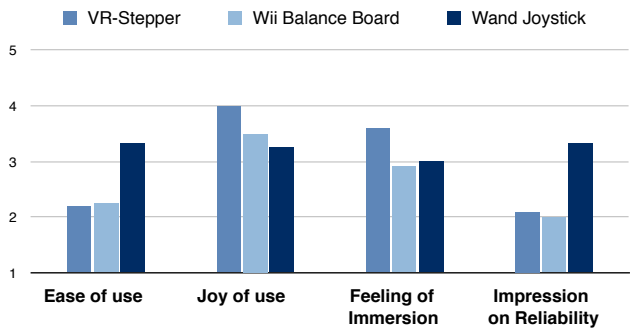

Figure 4. Quantitative results for each test criteria.

*Feeling of Immersion*
A one-way ANOVA did not yield any significant difference ($p > .05$) between devices that are averaging around 2.9-3.6.

*Impression on Reliability*
A one-way ANOVA found a significant difference in terms of reliability ($F_{2,18} = 22.18$; $p < .0001$). A Tukey HSD Test determined that the joystick ($M= 3.4$) was perceived as more reliable than the VR-Stepper ($M= 2.1$; $p < .01$) and the Wii Balance Board ($M= 2.0$; $p < .01$).

**Discussion**
Analyzing the data (**Figure 4**) showed that using a Wand Joystick is significantly easier. Confirming the hypothesis H3, the reliability was also significantly rated much higher. The VR-Stepper was also experienced, in terms of immersion and joy slightly better than the other tested interfaces, as expected. Unfortunately, the difference was not statistically significant, thus hypothesis H1 and H2 did not confirm. Therefore this short study could not prove that a physical movement of legs results in a raise of immersion. However a broader study with a doubled sample size and similar feedback would deliver a statistically significant difference. For future studies additionally measuring real quantitative data (e.g. task completion time) would also be interesting for a comparison of locomotion interfaces.

When being in a CAVE-like installation and using stationary IR tracking systems, the additional IR sensor attached to the Stepper might not operate due to the overshoot of great IR noise. In that case, crouching has to be detected by the local tracking system in a CAVE. The problem of losing balance also sometimes occurred in the CAVE, when the user was fully focusing on the game and *"forgetting about being on a stepper"* thus P4 also stated *"it can be dangerous by falling from it"*. Two users stated the stepper to be *"exhausting but very funny"*. P8 proposed to create a skiing-application. Except from one outlier we excluded (P11 - a female mid-age subject, who fall off the stepper and did not wish to complete the study) the overall feedback was quite good. Most users agreed that using the Wii Balance Board required an extra familiarization phase, which is not needed for the stepper. Overall, all study participants really enjoyed the user test, as they said during and after the test.

**CONCLUSION**
This paper demonstrated a novel approach of hacking a stepper as a low-cost DIY input interface for virtual environments. Furthermore a small study was conducted, where the functionality of this device was tested in a self-built 3D scene. Additionally a comparison to a "Wii Balance Board" and a "Wand Joystick" is accomplished, which gave valuable results such as: an instant / direct input action with a haptic feedback from the buttons by the Wand was preferred to use. Never the less, the VR-Stepper was rated better in terms of "joy of use" and "feeling of immersion", yet more tests need to be conducted in order to report empirical evidence. Overall, the observation of our study showed that very fast movements or lags in a 3D scene dramatically increase cyber sickness and lead to a loss of equilibrium, as we agree with Pausch [7] who already stated the importance of low latency HMD's. An advantage in VE's could include exercise programs, which even would provide significantly more joy of use [10].